\documentclass[12pt]{iopart}
\usepackage{epsf}
\begin{document}

\title[The R\^ole of Strangeness in Astrophysics]{The R\^ole of Strangeness in
Astrophysics --- an Odyssey through Strange Phases}

\author{J\"urgen Schaffner-Bielich}

\address{Department of Physics, Columbia University, 538 West 120th Street, New
York, NY 10027, USA}

\ead{schaffne@nt3.phys.columbia.edu}

\begin{abstract}
The equation of state for compact stars is reviewed with special emphasis on
the role of strange hadrons, strange dibaryons and strange quark
matter. Implications for the properties of compact stars are presented.  The
importance of neutron star data to constrain the properties of hypothetic
particles and the possible existence of exotic phases in dense matter is
outlined. We also discuss the growing interplay between astrophysics and
heavy-ion physics.
\end{abstract}




\section{Introduction}

In the last few years, it became clear that strangeness has to be included as
another degree of freedom in astrophysical systems. In this review, we will
outline some recent developments in the study of matter with strangeness under
extreme conditions with relevance to astrophysics.  The field is growing
rapidly, theoretically as well as experimentally, so we will not give an
overview of the field but rather focus on some recent works about the
properties of compact stars with a remark about the cosmological phase
transition including strangeness.

The topics we are going to cover are: hyperons in neutron stars, kaon
condensation in hadronic matter, H-dibaryon condensation, strange quark phase
and twins, kaon condensation in quark matter, signals from proto-neutron stars
with a strange phase, strange quark stars, new data from a isolated neutron
star and the cosmological phase transition with strangeness.

The phase diagram of Quantum Chromodynamics (QCD) can be studied at large
temperature and zero (or small) density by lattice calculations and by
relativistic heavy-ion collisions. Neutron stars on the other hand probe QCD
at high density and small temperature and provide therefore a complementary
laboratory to study QCD under extreme conditions. Created by supernova
explosions, neutron stars are compact remnants with masses around 1--2 solar
masses and radii of about $R\sim 10$ km. The central density of the compact
star will be then several times normal nuclear matter density $n_0$. More than
1000 pulsars, rotating neutron stars, are known today. The most precisely
measured mass is the Hulse-Taylor pulsar with $M=(1.4411\pm0.00035)M_\odot$.

\section{Strange hadrons in compact stars}

The first systematic theoretical investigation of the composition of a neutron
star was done in reference \cite{Ambart60}. Nucleons as well as hyperons were
included in the equation of state, nevertheless, in this early work they were
treated as free particles. Hyperons were found to appear at $4 n_0$
($\Sigma^-$) and $8n_0$ ($\Lambda$'s). The corresponding neutron star have too
small a maximum mass of only $M_{max}\sim 0.7M_\odot$ to agree with the
observed pulsar masses. Therefore, strong interactions need to be incorporated
for describing neutron stars! Investigations in different effective models in
the last few years now essentially agree on a critical density of $n_c\sim 2
n_0$ (see \cite{Glen85}) for hyperons to appear in neutron star matter:
effective nonrelativistic potential models \cite{Balberg97}, the Quark-Meson
Coupling Model \cite{Pal99}, extended Relativistic Mean-Field approaches
\cite{Knorren95b,SM96}, Relativistic Hartree-Fock \cite{Huber98},
Brueckner-Bethe-Goldstone \cite{Baldo00}, Brueckner-Hartree-Fock
\cite{Vidana00}, Density-Dependent Hadron Field Theory \cite{Hofmann2000} and
chiral effective Lagrangians \cite{Hanauske00}.

The abundance of hyperons in the interior raises the question about effects
from hyperon-hyperon interactions which are (besides the known double
hypernuclear events) essentially unknown. Hyperon-hyperon interactions can
have drastic effects even on the global properties of neutron stars
\cite{SHSG00}: a phase transition to hyperonic matter is possible which
generates a third family of solution of compact stars besides white dwarfs and
neutron stars. Even self-bound compact stars can be generated for large
attraction with rather small radii of about $R\sim 7-8$ km. These hyperstars
are also of relevance for relativistic heavy-ion collisions. Strange hadronic
matter with a strongly attractive hyperon-hyperon interaction produces a
second minimum at finite strangeness which is in accord with hypernuclear
data.  Matter in that second minimum is long-lived but not absolutely stable
as there is an energy barrier only allowing multiple weak decays but not
single weak decays \cite{SHSG00}.

Kaons can also appear in dense hadronic matter (for quark matter see the next
section). As they have zero spin, they will form a Bose condensate in a
neutron star (for a recent review see \cite{RSW01}). For a sufficiently
reduced effective energy of the antikaons, neutrons will be transformed to
protons and antikaons or equivalently electrons to $K^-$ and neutrinos.  The
appearance of a kaon condensed phase depends crucially on the $K^-$ optical
potential at $n_0$ which can be estimated from coupled channel calculations
\cite{Koch94, Waas97} ($-120$ MeV), selfconsistent calculations
\cite{Lutz98,Ramos00,SKE00,Tolos01} ($-30$ to $-80$ MeV) or $K^-$ atomic data
\cite{Fried94,Fried98} ($-180$ MeV) . A recent combined chiral analysis of the
available data finds a rather shallow potential of only $-55$ MeV
\cite{Cieply01}. The impacts for the mass-radius relation can be drastic if
the optical potential for the $K^-$ is larger than about $-120$ MeV: compact
stars with a kaon condensate can have radii of only $R\sim 8$ km
\cite{GS98L,GS99} and produce again a third family of solution when taking
into account the $\bar K^0$ also \cite{Banik01}.  A large region in the
interior of the star may form geometric structures in the mixed phase of
nuclear matter and kaon condensed matter
\cite{GS98L,GS99,Reddy2000} which are determined by
one model Lagrangian describing both phases simultaneously. The diameter of
these structures is about 10--30 fm, and the spacing about 5--10 fm.

Other exotic particles can also appear inside a neutron star, like dibaryons
or strangelets. Their appearance in dense matter marks the onset of the mixed
phase to the deconfined quark matter. The H-dibaryon is a hypothetical
six-quark state which is antisymmetric in flavor, colour and spin, hence it is
also a boson with zero spin \cite{Jaffe77}. The stability of the H-dibaryon is
closely connected with double $\Lambda$ hypernuclear data as two $\Lambda$'s
have the same quark content as the H-dibaryon and can be transformed into it
via strong interactions. In addition to the three 'old' double hypernuclear
events, two new events were reported this year: a $^{~~4}_{\Lambda\Lambda}$H
by experiment E906 at BNL, and $_{\Lambda\Lambda}^{~~6}$He by experiment E373
at KEK (for references and a detailed interpretation of the presently
available data see \cite{FG01}).  The double $\Lambda$ hypernuclear data
suggest an attractive force between the two $\Lambda$'s. No strong decay to
the H-dibaryon has been observed which would give a lower limit to the mass of
the H-dibaryon of $m_H> 2m_\Lambda-B_{\Lambda\Lambda}\sim 2220 MeV$. Here
$B_{\Lambda\Lambda}$ is the binding energy of the two $\Lambda$'s. If the
H-dibaryon appears in dense matter it will form a Bose condensate. The
presence of the H-dibaryon will shift the maximum mass of a neutron star down
but the corresponding radius up.  This effect provides a constraint on the
property of the H-dibaryon in dense matter as the maximum mass can not be
lower than $M=1.44M_\odot$. A deeply bound H-dibaryon with an attractive
nuclear potential turns out to be not compatible with this constraint
\cite{GS98H}.

\section{Strange quarks in compact stars}
\label{sec:quarks}

At sufficiently high density, the transition to deconfined strange quark
matter should appear which is mostly modeled by the MIT bag model. It has
been pointed out only in the last year, that this transition allows for the
existence of a new class of compact stars: strange quark star twins
\cite{GK2000,Schertler00}. They are called twins, because there are two
solutions to the Tolman-Oppenheimer-Volkov equations which have similar masses
but different radii, one without and the other with a pure quark core. The
compact stars with a quark core constitute a third family of solutions besides
white dwarfs and neutron stars. The solution is stable and can in principle
appear for (any) first order phase transition as pointed out already by
Gerlach in 1968 \cite{Gerlach68}. The twin star solution within the MIT bag
model gives radii which are only slightly smaller than for the neutron star
solution, $R=10$--12 km, and appear only for a narrow range of the bag
constant, $B^{1/4}=178$--182 MeV \cite{Schertler00}. We point out, that such
radii can be also reached without having a third solution. The unambiguous
signal for twin stars would be the determination of the mass and radii of
neutron stars which have similar masses but different radii.

The detailed properties of the quark phase in compact stars has been a topic
of recent interest (for a review see \cite{ABR01}). Here, we just report on
one of the most recent developments in this rapidly evolving field of color
superconductivity concerning strangeness. For three-flavour QCD with massless
quarks, a colour-flavour locked (CFL) phase will form. For a finite strange
quark mass, the CFL phase will be under stress. Low lying excitations appear
which have the quantum numbers of the pseudoscalar meson octet, i.e. of pions
and kaons. These pionic and kaonic excitations can form a pion or kaon
condensate in the quark phase of the compact star
\cite{Schafer00}. It is interesting to note that kaon condensation
in quark matter actually reduces the number of strange quarks contrary to kaon
condensation in hadronic matter.  The kaon condensed phase in quark matter
will not change the global properties of the compact star as the effects on
the equation of state are likely to be negligible. But the transport
properties of quark matter will be affected strongly.

\section{Signals for a strange phase in compact stars}

There are at least two signals for the presence of strange hadrons and/or
strange quarks in the interior of neutron stars which are common for all
strange phases regardless of the details of the composition. We discussed one
common signal so far, namely that a third family of compact stars can appear
with smaller radii than ordinary neutron stars which has been shown to be the
case for hyperon matter \cite{SHSG00}, for kaon condensation \cite{Banik01}
and for strange quark matter \cite{GK2000}.

Another signal has been proposed recently which again is possible for the all
three kinds of strange phases: the neutrino flux from a proto-neutron star in
a supernova \cite{Pons01L}. A newly born hot neutron star with a strange phase
can support more mass than a cold one so that the cooled neutron star has to
collapse to a black hole. The flux of neutrinos emitted from the supernova
event will stop suddenly when the black hole is formed. The delayed collapse
happens again for every strange phase discussed so far, be it with hyperons,
kaons or strange quarks. The timescale to the instability lies typically below
a minute and is slightly different depending on the critical density at which
strange particles appear.  The mass range of the instability is rather
similar, $M=(1.6$--$2.2)M_\odot$, so low mass black holes will be produced in
either scenario.

\section{New approaches to dense QCD}

The 'standard' model for dense QCD was to use the MIT bag model with free
quarks, whose results depend strongly on the unknown bag parameter.  Only
recently did one start to explore other approaches to dense QCD, all but the
first includes effects from the strange quarks: Schwinger-Dyson model
\cite{Blaschke99}, massive quasiparticles
\cite{Schertler98,Peshier00}, Nambu--Jona-Lasinio model
\cite{Schertler99} and perturbative QCD
\cite{FPS01,FPS01b}. We will focus on the latter one in the following in more
detail. We note that we discuss in this section the global properties of
bulk quark matter, while section \ref{sec:quarks} about color
superconductivity deals with two quark interactions. 
Both pictures are then to be seen complementary.

The underlying new physical picture for cold and dense QCD is that the phase
transition is from hadrons/massive quarks to massless quarks, i.e.\ one has to
study the chiral phase transition not the deconfinement transition. The
transition from hadrons to massive quarks is smooth as there exists no order
parameter at zero temperature and finite density contrary to the chiral phase
transition. We take the thermodynamic potential up to second order in the
strong coupling constant $\alpha_s$ 
as a model for dense QCD and compute
thermodynamically consistent the pressure, number density and energy
density. The coupling constant $\alpha_s$ is running and depends on the
renormalization subtraction point $\bar \Lambda$.

The behaviour of the perturbative series looks reasonable, but the results
depend so strongly on $\bar \Lambda$ that it is constrained by physics with a
reasonable range of $\bar \Lambda/\mu=2$--3 (case 2 and 3 in the following)
where $\mu$ is the quarkchemical potential.  Note that different values of
$\bar \Lambda$ correspond to different scales, i.e.\ different chemical
potentials. As the density grows like $\mu^3$ a slightly smaller scale results
in a tremendous increase in the density.  In every case, the pressure turns
out to be far from a free gas even for large values of the quark chemical
potential. There is a 30\% reduction in the pressure compared to a free gas at
$\mu=1$ GeV, and still a 7\% reduction at $\mu=100$ GeV! So quark matter can
definitely not be described by a free gas, especially not for densities of
interest for compact stars. For all cases does the pressure vanish at some
$\mu_c$ (here for certain does perturbative QCD break down). The maximum mass
for a pure quark star, ignoring for a moment a hadronic mantle, is
$M=1.05M_\odot$ for case 2 and $M=2.14M_\odot$ for case 3 with radii of
$R=5.8$ km and $R=12$ km, respectively \cite{FPS01}.
Similarly small values for the mass and radius of quark stars were also
found in \cite{Peshier00}. 

The matching to a low density equation of state can be classified to be either
weak or strong. For a weakly first order phase transition or a crossover, the
pressure rises strongly before it reaches the pressure curve for the massless
quarks. For a strong first order transition, the pressure has to rise
slowly. Interestingly, asymmetric matter up to say $2n_0$ can be parameterized
by $E/A\sim 15$ MeV $n/n_0$ (see \cite{Akmal98}), so that the baryonic
pressure is just 4\% that for a free quark gas at $n_0$, $p_B\sim 0.04
(n/n_0)^2 p_{\rm free}$! This slow rise of the pressure at least allows for a
strong first order transition. Adding a hadronic equation of state to the
quark equation of state, gives ordinary neutron stars with a quark core
(hybrid stars) for both cases. In addition, a strong phase transition enables
again a third class of compact stars, this time with masses of only $M\sim
1M_\odot$ and radii of only $R\sim 6$ km! The quark phase dominates the
composition of the compact star, as the maximum density in the quark core is
$15n_0$ and the hadronic pressure is small compared to the quark pressure
\cite{FPS01b}. 

The mass-radius relation of neutron stars can therefore tell us something
about the order of the phase transition in dense QCD. The analysis of the data
from isolated neutron stars will be able to pin down the mass and the radius
simultaneously. The radio-quit neutron star RX J185635-3754 is the closest one
known to our planet \cite{Walter2001}. The spectra is nearly thermal with an
effective temperature of $T\sim49$ eV. First parallax measurements by the
Hubble Space Telescope with a distance of about 60 parsec give a black-body
radius of only $R_\infty=6$ km \cite{Walter2001}. A more detailed analysis of
the combined data from ROSAT, EUVE and HST, including effects from the
atmosphere, finds a best-fit mass and radius of $M\sim0.9M_\odot$ and $R\sim6$
km \cite{Pons2001}! The authors conclude that these values together are not
permitted by any plausible equation of state, even not that for absolutely
stable strange stars. We note, however, how close they are to the above values
for quark stars within perturbative QCD. Nevertheless, the parallax
measurement in \cite{Walter2001} is questioned in \cite{KKA01} and a
radius of $R_\infty=15\pm6$ km is found. The release of
the newest data from HST next year will settle this discrepancy.
Gravitational micro-lensing is another promising tool to find small compact
stars. The MACHO project reports events with a mass of $M\sim 0.5M_\odot$ from
several years of observation towards the Large Magellanic Cloud
\cite{MACHO}. Events with $M=0.13M_\odot$ and $M=0.25 M_{\rm Jupiter}$ are
detected at the globular cluster M22 \cite{M22}.

\section{Strange cosmological phase transition}

Strangeness is also of interest for the early universe, as strange stars might
have been produced there. The newly formed strange star faces several
obstacles to survive (see \cite{Madsen99o} for an overview): the mass number
must be smaller than $A<10^{49}$ due to the horizon at a time of $t\sim 1\mu
s$ after the big bang, $A>10^{23}$ so that the neutron capture rate is not in
conflict with the big bang nucleosynthesis and $A>10^{30-40}$ so that the
strange star does not evaporate completely at a temperature of $T\sim 160$
MeV. A typical compact star on the other hand has $A\sim 10^{57}$. A possible
solution is to propose that an inflation happens at the quark-hadron phase
transition \cite{Borghini00}. The exponential expansion and supercooling
allows the production of large quark stars with masses of
$M\sim(10^{-2}-10)M_\odot$. This scenario is interesting as it is able to
produce small compact stars below $1M_\odot$. Standard supernova simulations
usually get neutron stars above that mass. The crucial point for quark
inflation to happen is to assume a large surface tension of $\sigma=50-120$
MeV fm$^{-2}$. Therefore, hadron bubbles have to expand tremendously to
overcome the surface energy which causes the inflation.  Note that such a
large surface tension would make the mixed phase region from hadrons to quarks
in compact stars disappear!

\section{Astrophysics versus Heavy-Ion Physics}

There are strong relations between the field of astrophysics and heavy-ion
physics. Following the topics discussed here, we list some of them.

Hyperstars, compact stars composed of strongly attractive hyperonic
matter, are connected to strange hadronic matter and MEMOs
(see \cite{SG00} and references therein). If hyperstars 
exist, strange hadronic matter has to exist and can possibly be formed in
relativistic heavy-ion collisions. 

Kaons in the medium can be studied as well as for kaon condensation in neutron
stars as well as for the subthreshold production of antikaons in heavy-ion
collisions. If kaon condensation is so strong as to change the global
properties of compact stars, it certainly will impact the production of
antikaons in dense matter. Nevertheless, the optical potential probed in
heavy-ion collisions will be in general different from that in neutron stars
so that a direct comparison has to be taken with care (see \cite{SKE00})!

Properties of the H-dibaryon in the medium as well as other exotic stable
particles, which dedicated heavy-ion experiments are hunting for,
can be constrained by astrophysical (pulsar) data. Dibaryons with strangeness
can be formed in relativistic heavy-ion collisions and it is possible to detect
them by their decay properties \cite{SMS00}.

The existence of strange stars is ultimately connected with the stability of
strangelets and hence, their searches in relativistic heavy-ion collisions,
too (see \cite{Klingen99} and references therein). If strange stars exist,
strangelets accessible to heavy-ion experiments are likely to be stable but
they are short-lived when taking into account shell effects (for an overview
see \cite{Greiner99o}).

Quark matter at finite density behaves highly nonideal around the chiral phase
transition similar to matter at finite temperature around $T\sim T_c$ (see
e.g.\ \cite{Andersen99}) which is probed in relativistic heavy-ion collisions.

Finally, the cosmological phase transition might allow for an inflation.  If
this is the case, then it will have strong impacts on the study of the
deconfinement phase transition in ultrarelativistic heavy-ion collisions and
can result in an explosion. The recently measured exotic two-particle
correlation at the Relativistic Heavy-Ion Collider might be an indicative for
such a scenario \cite{Dumitru01}.

\ack 
We thank 
Eduardo Fraga, Norman Glendenning, Matthias Hanauske and Rob Pisarski
for fruitful discussions and 
Brookhaven National Laboratory for their kind hospitality where parts
of this work were completed.  We acknowledge support from the DOE
Research Grant, Contract No.\ DE-FG-02-93ER-40764.

\section*{References}


\end{document}